# Molecular Beam Epitaxy of Lithium Niobium Oxide Multifunctional Materials

M. Brooks Tellekamp, Joshua C. Shank, and W. Alan Doolittle[*]

*Georgia Institute of Technology, 791 Atlantic Drive Atlanta, Ga 30332*

[*]Corresponding email: alan.doolittle@ece.gatech.edu

*Abstract—* **The role of stoichiometry and growth temperature in the preferential nucleation of material phases in the Li-Nb-O family are explored yielding an empirical growth phase diagram. It is shown that while single parameter variation often produces multi-phase films, combining substrate temperature control with the previously published lithium flux limited growth allows the repeatable growth of high quality single crystalline films of many different oxide phases. Higher temperatures (800-1050°C) than normally used in MBE were necessary to achieve high quality materials. At these temperatures the desorption of surface species is shown to play an important role in film composition. Using this method single phase films of NbO, $NbO_2$, $LiNbO_2$, $Li_3NbO_4$, $LiNbO_3$, and $LiNb_3O_8$ have been achieved in the same growth system, all on c-plane sapphire. Finally, the future of these films in functional oxide heterostructures is briefly discussed.**

## 1. INTRODUCTION

The lithium niobium oxide family consists of conducting, semiconducting, and insulating materials across a wide resistivity and bandgap range. Figure 1 lists the resistivity of a few materials in the system as a function of niobium valence, ranging from conducting niobium to insulating lithium niobite ($LiNbO_3$). These materials span 22 orders of magnitude in resistivity with bandgaps from IR to UV [1–8]. Oxides in general have many desirable multifunctional properties, for example; piezoelectric, pyroelectric, and ferroelectric effects which can exist in a single material such as lithium niobate [9,10]. Lithium niobite ($LiNbO_2$), a suboxide of the same family, is currently the focus of multiple research areas. $LiNbO_2$ is used as a memristor for neuromorphic applications, a battery cathode material showing potential for high rate capability and long term cycle stability, and is also studied for unique optical properties [11–13]. $NbO_2$ acts as a digital memristor, a device with discrete on and off resistance states. $NbO_2$ is currently used in memory, neuristor circuitry, and relaxation oscillator circuitry [14–16]. The Li-Nb-O material family also includes other ceramics of various dielectric constants used in a variety of applications including battery electrodes, microwave frequency dielectrics, phosphors, photocatalysts for water reduction, and hysteretic MIM tunnel diodes or "memdiodes" ($Li_3NbO_4$, $LiNb_3O_8$, and $Nb_2O_5$) [17–22,5,23,24].

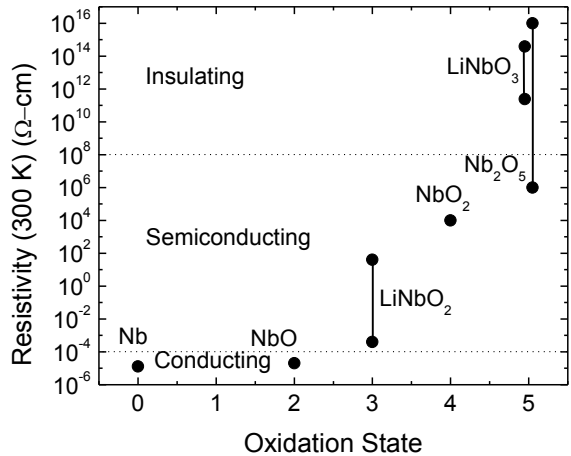

**Figure 1.** (Lithium) Niobium oxides plotted by their resistivity versus niobium valence, spanning 22 orders of magnitude resistivity. Materials can be conducting, semiconducting, or insulating depending on the oxidation state. The insulators $Li_2O$, $Li_3NbO_4$, and $LiNb_3O_8$ are omitted for clarity.

The growth optimization of these materials in a single deposition system is desirable due to the variety of properties that can be achieved and the potential heterostructure devices that can be imagined. Among these possibilities are strain enhanced sensors and MEMS, high K dielectrics, tunable dielectrics, superlattices, thin film batteries, memtransistors, multi-layer neuromorphic heterostructures, and ferroelectric switching transistors with switchable enhanced channel conductance [25,26].

Despite the potential for new and enhanced functional devices and materials, this Li-Nb-O multifunctional material family remains relatively unexplored. A many-phased oxide system inherently contains significant complexity. The phase diagram of the $Li_2O$-$Nb_2O_5$ system is complicated by multiple phases existing in a very small range of temperature and stoichiometry [27]. Due to these tight tolerances, precise control and understanding of all parameters is necessary for controllable growth of heterostructures and single phase materials. Molecular Beam Epitaxy (MBE) is an excellent platform for high stoichiometry control and crystalline quality, and therefore a good candidate for controlled growth of these materials [25]. The MBE growth of Li-Nb-O materials using $NbCl_5$ as a high vapor pressure niobium precursor has been previously described by a lithium controlled niobium incorporation rate method [28]. Briefly, $NbCl_5$ can be reduced to elemental niobium in the presence of lithium and a hot substrate where lithium getters the chlorine. LiCl desorbs due to its high vapor pressure leaving elemental niobium to react with oxygen and excess lithium. In the same work a method of single parameter phase control in the Li-Nb-O system is described where the lithium supply is limited while niobium and oxygen are supplied in excess. The lithium flux directly controls the amount of adsorbed niobium so that adjusting the Li flux modifies niobium to oxygen ratio, and therefore the oxidation state, with respect to the constant $O_2$. This study was performed at a single substrate temperature, 900°C, and in practice often produces multiple phases laterally across a film due to temperature gradients across the surface at high temperature growths. In this work it is shown substrate temperature plays a role in the preferential nucleation of various phases in the Li-Nb-O family, and that by combining this new understanding with the above lithium controlled niobium incorporation rate method single phase films spanning the entire phase space can be epitaxially grown.

## 2. EXPERIMENT

### 2.1 Growth System Modifications

The materials in this work are grown in a Varian Gen-II style MBE system which has been customized for the growth of lithium niobium oxides using a lithium assisted metal-halide growth chemistry discussed in detail by *Henderson et. al* [28]. The method involves the sublimation of $NbCl_5$ from a custom water and filament heated near-ambient cell, described elsewhere, allowing controllable evaporation of high vapor pressure materials (typical bulk operating temperatures are 35 – 50°C) [29].

A modified Veeco corrosive series antimony cracker with a custom solid tantalum crucible is used to supply lithium to the system where it getters chlorine from $NbCl_5$ and HCl. Excess lithium may be incorporated into the film, and along with the niobium in the presence of oxygen can form a variety of phases in the lithium-niobium-oxygen phase system such as $LiNbO_2$, $LiNbO_3$, $Li_3NbO_4$, and $LiNb_3O_8$. The phase which is grown depends on the Li:$NbCl_5$ flux ratio, the supplied oxygen, and the substrate temperature as will be shown in this work.

A custom Vesco-NM substrate heater is used to allow growth at high temperatures in a corrosive oxygen and chlorine environment. The heater consists of a thick tantalum filament fully encased in PBN, which can be heated to 1000°C in an oxygen background of $5 \times 10^{-5}$ torr.

### 2.2 Substrate Preparation

All films in this work are grown on c-plane (006) oriented sapphire substrates. The substrates are cleaned by acetone, methanol, and isopropyl alcohol followed by 4:1 $H_2SO_4$:$H_2O_2$ at 120°C before loading and degassing in a separate chamber, being careful during loading to minimize sample contact with the surrounding metal substrate holder which will conduct thermal energy away from the sample creating thermal non-uniformity.

### 2.3 Characterization

Films were structurally characterized by a Phillips Xpert MRD diffraction system using Cu-$K_\alpha$ x-rays ($\lambda$ = 1.54056 Å). Phase identification of lithium niobium oxides by x-ray diffraction (XRD) is challenging due to the existence of multiple phases with similar crystal structures. There are five possible phases in the narrow range 36.772 – 37.069 degrees 2θ-ω, and 3 of those materials fall within 0.4 degrees of each other, making phase identification extremely difficult. It is possible to differentiate them all, however, using multiple XRD scans as well as knowledge of the material in question. In this work film transparency, the existence of double or half x-ray reflections, and asymmetric diffraction scans are analyzed together to definitively identify material phases [30]. The surface morphology of the films was characterized by a Veeco atomic force microscopy (AFM) system.

## 3. RESULTS AND DISCUSSION

### 3.1 Growth Parameters

Previous work in Li–NbCl$_5$–O$_2$ growth chemistry indicated that phase control was best mediated by a lithium controlled niobium incorporation rate approach. Because the niobium supply is controlled by lithium, the niobium to oxygen ratio can be adjusted by varying a single parameter, the Li flux. It was shown that through this method increasing the Li flux produced films of reduced oxidation state, resulting in a progression from Li$_3$NbO$_4$ → LiNbO$_3$ → LiNbO$_2$ → NbO as the Li flux is increased holding all other fluxes constant at constant substrate temperature [28].

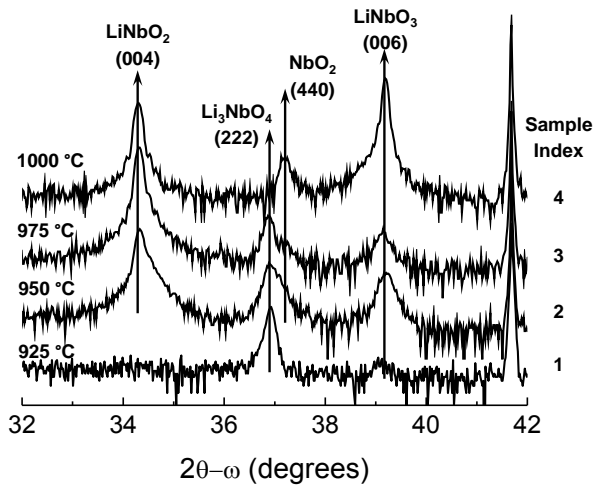

**Figure 2.** X-ray diffractograms of films grown at increasing substrate temperature while holding all fluxes constant ($\phi = 0.6$). From bottom to top in 25 degree increments, 925°C primarily produces Li$_3$NbO$_4$, 950 and 975°C primarily produces LiNbO$_2$, and 1000°C primarily produces LiNbO$_3$. All films are multi-phase. The temperature dependence is attributed to increased Li desorption.

Motivated by the prevalence of multi-phase films, temperature studies were carried out to analyze the effect of substrate temperature on the nucleation of various phases. Figure 2 shows the x-ray diffractograms of films grown at constant flux while substrate temperature is increased from 925 - 1000°C in steps of 25 degrees. For these growths the beam equivalent pressures as measured by Bayard-Alpert ion flux gauge were approximately $6 \times 10^{-7}$ torr for Li and $1 \times 10^{-6}$ torr for NbCl$_5$. Further data will be presented as Li:NbCl$_5$ flux ratio, with all data presented for growths at the same O$_2$ flow, where $\phi = \phi_{Li}/\phi_{NbCl5}$ and $\phi$ is source flux measured in beam equivalent pressure (BEP), in units of torr. The O$_2$ flow was set at 4 SCCM by mass flow controller (MFC) resulting in a background oxygen pressure of approximately $4.5 \times 10^{-5}$ torr. The 925°C substrate temperature condition resulted in (222) oriented Li$_3$NbO$_4$ with a small fraction of c-oriented LiNbO$_3$. Increasing the substrate temperature to 950°C resulted in c-oriented LiNbO$_2$ along with Li$_3$NbO$_4$, (440) NbO$_2$, and LiNbO$_3$. Further increasing the substrate temperature to 975°C produced higher quality LiNbO$_2$ as evidenced by Pendellösung fringes and an decrease in (002) XRD rocking curve from 410 to 307 arcseconds. Li$_3$NbO$_4$, NbO$_2$, and LiNbO$_3$ are also present in the film. Finally, increasing the substrate temperature to 1000°C significantly increased the LiNbO$_3$ diffraction signal with respect to all others, and the Li$_3$NbO$_4$ phase was eliminated.

Further investigation reveals that the optimal growth temperature and $\phi$ are coupled parameters for obtaining single phase material. Figure 3 shows the effect of $\phi$ on phase nucleation at 975 °C. At a Li-poor flux of $\phi = 0.1$ there is not enough Li to incorporate into the growing film, and at 975 °C pure phase $NbO_2$ nucleates. Increasing $\phi$ to 0.8 results in the appearance of $LiNbO_2$ and $LiNbO_3$ and the reduction of $NbO_2$ to a small fraction of the film indicating the presence of a multi-phase region in $\phi$. Further increase to unity $\phi$ results in pure phase $LiNbO_2$, while additional Li at $\phi = 2$ introduces the Li-rich phase $Li_3NbO_4$.

Figure 4 explores the same trend at 1000 °C. A Li-poor flux ratio of 0.2 still results in $NbO_2$, but the higher substrate temperature allows small fractions of $LiNbO_3$ and its Li-poor counterpart $LiNb_3O_8$ to form. Increasing the flux ratio to 0.4 prevents $NbO_2$ from forming; however there is still not enough Li for pure $LiNbO_3$ and the Li-poor phase ($LiNb_3O_8$) is still present. Finally, increasing $\phi$ to 0.8 results in pure $LiNbO_3$ in contrast with the 975 °C case where the same flux ratio produces both $LiNbO_3$ and $LiNbO_2$.

With this in mind, optimal temperature ranges can be approximately constructed based on experimental data at various flux ratios. Table 1 presents the temperature ranges appropriate for the growth of various material phases at 4 SCCM $O_2$ flow.

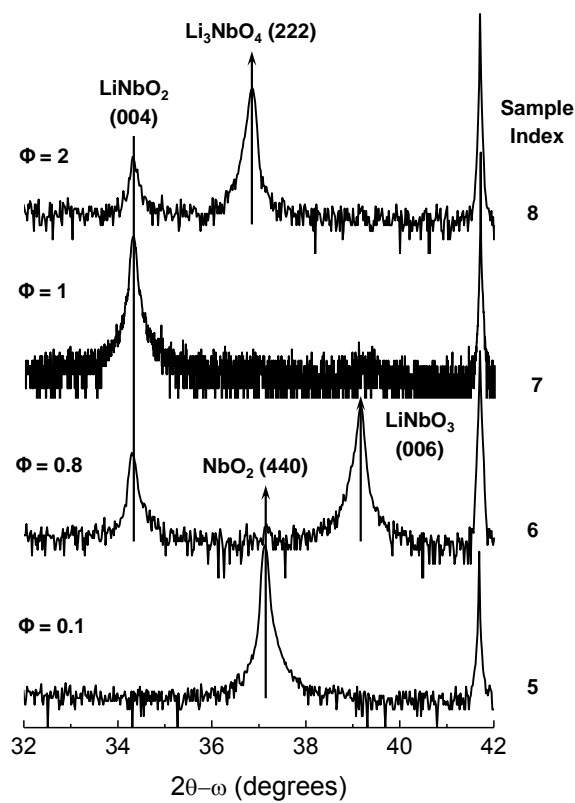

**Figure 3.** X-ray diffractograms of films grown at 975 °C and increasing $\phi$ from bottom to top. As $\phi$ increases the films transition from $NbO_2$, through $LiNbO_3$ and $LiNbO_2$, to $Li_3NbO_4$.

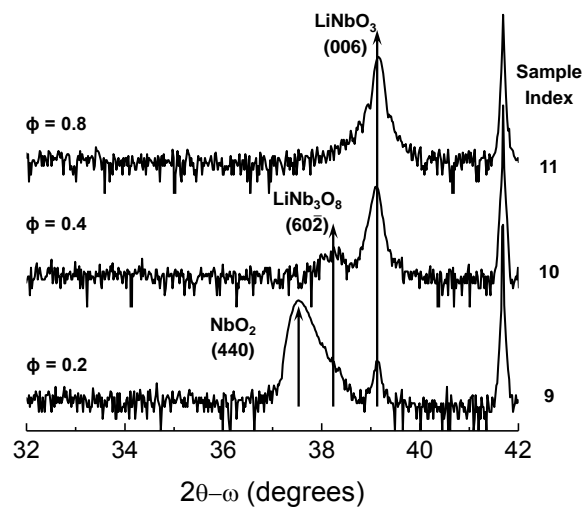

**Figure 4.** X-ray diffractograms of films grown at 1000 °C and increasing $\phi$ from bottom to top. As $\phi$ increases the phase progresses from $NbO_2$, through $LiNb_3O_8$, to $LiNbO_3$.

**Table 1.** Optimal substrate temperatures for MBE growth of Li-Nb-O phases at 4 SCCM $O_2$ flow.

| Material | Optimal Temperature (°C) |
|---|---|
| Nb | > 850 |
| NbO | 800 – 900 |
| $NbO_2$ | 950 – 1000[†] |
| $LiNbO_2$ | 950 – 975 |
| $LiNbO_3$ | 1000 – 1050[*] |
| $Li_3NbO_4$ | 850 – 925 |
| $LiNb_3O_8$ | 975 – 1050[*,†] |

[*]1050°C is the upper operational limit of the substrate heater. The optimal range may extend beyond this limit.
[†]Further optimization not pursued.

Using optimized substrate temperatures allows more control when employing the lithium controlled niobium incorporation rate method. When growing at optimal temperatures, flux conditions can be slightly altered to promote single phase growth. Temperature conditions from Table 1 were applied to the Li-limited growth technique, increasing ϕ for materials of lower oxidation state and decreasing it for higher oxidation state, to achieve single phase films of different phases. Figure 5 shows symmetric x-ray diffractograms of phases grown using this combined method.

$LiNbO_2$, NbO, $Li_3NbO_4$, $LiNbO_3$, $LiNb_3O_8$ and $NbO_2$ are all grown single phase as evidenced by XRD. All of the films shown in Figure 5 are grown on c-plane sapphire by adjusting the $Li:NbCl_5$ flux ratio while growing in the optimal temperature regime defined in Table 1.

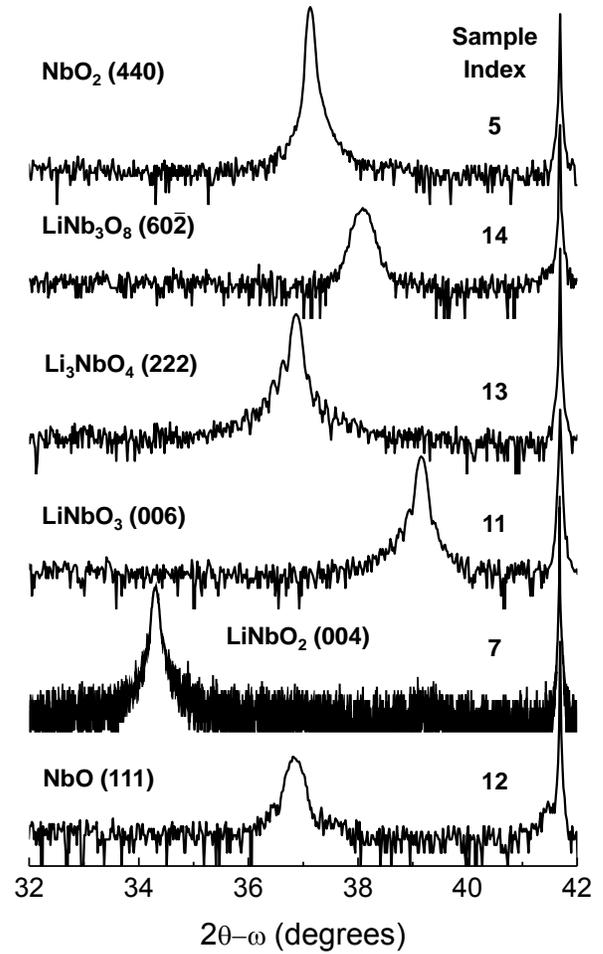

**Figure 5.** From bottom to top, c-oriented $LiNbO_2$, (111) oriented NbO, (222) oriented $Li_3NbO_4$, c-oriented $LiNbO_3$, (60$\bar{2}$) oriented $LiNb_3O_8$, and (440) oriented $NbO_2$, all grown single phase on c-oriented $Al_2O_3$.

*3.2 Phase Diagram Analysis*

Using the understanding outlined in section 3.1 and empirical results which followed this understanding a phenomenological phase diagram has been developed to represent the coupled effect of substrate temperature and flux ratio on the nucleation of niobium oxides and lithium niobium oxides. This diagram is shown in Figure 6 as a function of $Li:NbCl_5$ flux ratio, ϕ, and temperature, T. The boundaries and areas are phenomenologically and empirically derived based on growth parameters, XRD phase analysis, and optical transmission analysis. Figure 6 is not a true phase diagram in the classical sense but is intended as a visual guide to understanding the complex phase nucleation of niobium oxides and lithium niobium oxides on c-plane sapphire at high temperatures in high vacuum conditions. Figure 6 represents a number of key points, namely:

(1) Films grown below ~ 925 °C will be $Li_3NbO_4$ or NbO, depending on $\phi$,

(2) There exists a negative slope in the $\phi$-T space between region I and V due to the increase in $Nb:O_2$ ratio as $\phi$ is increased,

(3) There exists a positive slope in the $\phi$-T space at the II-III and III-IV boundary due to increased Li desorption at higher temperatures,

(4) There exists a mixed phase region from T ≈ 925 – 1000 °C and $\phi \approx 0.3 - 0.8$ represented in Figure 6 as a hashed oval, and

(5) Narrow growth windows exist for the desired phases III, IV, and V both in T and $\phi$.

As stated above, there is a mixed phase region in the middle of the diagram marked by a hashed oval. Growths in this region of $\phi$-T space result in multi-phase films which contain at least 2 but more often 3 phases. A detailed understanding of this region is not necessary because it is an undesired result, and is therefore not pursued. Additionally, multi-phase mixing is expected at the boundaries between phases though this mixed-phase result is not depicted. Finally, Figure 6 is presented in terms of a directly measurable quantity, flux ratio, rather than an absolute value. This allows the adoption of different $O_2$ flow rates by adjusting the overall cation flux values while maintaining the same ratio to oxygen anions. For the data given $\phi = 1$ represents $\phi_{Li} = 7 \pm 1.4 \times 10^{-7}$ torr BEP at 4 SCCM $O_2$ which results in ~ $4.5 \times 10^{-5}$ torr oxygen. Adjustment of either the $O_2$ flow rate or the absolute flux values individually will result in a modification of the growth windows, and can be visualized as a third orthogonal dimension to the diagram into and out of the page.

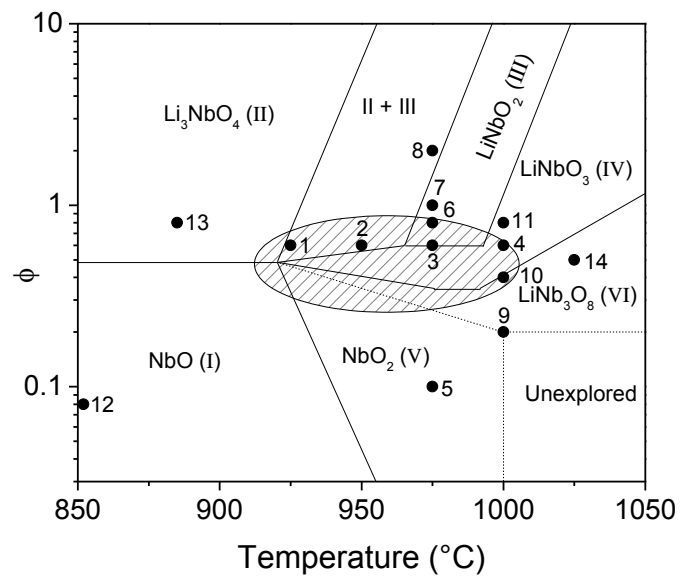

**Figure 6.** Phenomenologically and empirically derived phase diagram for the nucleation of niobium oxides and lithium niobium oxides on c-plane sapphire at high temperature and vacuum conditions. Data point indices correspond to the diffractograms labeled in Figure 2 - Figure **5**.

The oxygen deficient vacuum conditions in MBE, far from equilibrium, enable meta-stable sub-oxide phases like $LiNbO_2$ not possible at higher oxygen partial pressures. In the traditional phase diagram of the $Li_2O$-$Nb_2O_5$ system, recreated in Figure 7 from *Kitamura et. al.* [31] and *Reisman et. al* [27], $LiNbO_2$ is not present because sub-oxides are not possible in this fully oxidized system. Similarities may however be noted between the two diagrams. $Li_3NbO_4$ lies to the Li-rich side of $LiNbO_3$ in both diagrams while $LiNb_3O_8$ lies to the Li-poor side of $LiNbO_3$ in both diagrams existing at higher temperatures than $Li_3NbO_4$. NbO and $NbO_2$ are not present in Figure 7 because they do not contain Li, but can be grown by MBE due to the Li-poor conditions under which they are grown where all Li is desorbed as LiCl.

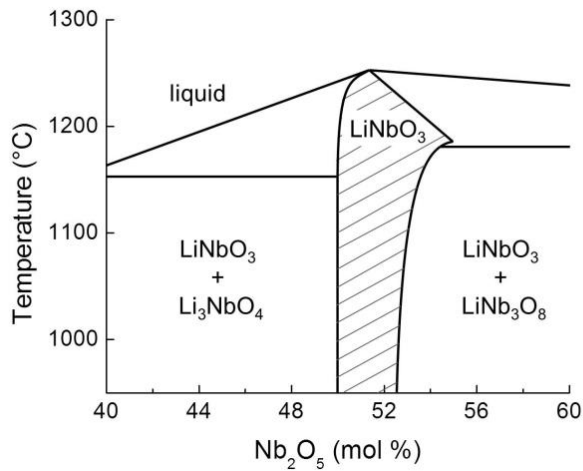

**Figure 7.** High temperature portion of the $Li_2O – Nb_2O_5$ phase diagram from 40 – 60 mol % $Nb_2O_5$. $LiNbO_2$ is not present because sub-oxides are not possible in this system. Analogous to Figure 6, $Li_3NbO_4$ appears to the Li-rich side of $LiNbO_3$ and $LiNb_3O_8$ appears to the Li-poor side and at higher temperatures. Figure redrawn from [27,31].

*3.3 Structural Characterization and Surface Morphology*

Structural and morphological characterization of the single phase materials was performed by XRD and AFM, respectively, to analyze the crystal quality of the (lithium) niobium oxides. The results of both methods are summarized in Table 2, showing symmetric rocking curves measured by XRD and RMS roughness measured by AFM. The symmetric rocking curve is a measure of film uniformity in the plane parallel to the substrate c-plane with higher values indicating increasing amounts of tilt with respect to that plane. RMS roughness quantifies the average height variation from the mean over a 1 $\mu m^2$ area.

Nb, $LiNbO_2$, $LiNbO_3$, and $Li_3NbO_4$ all show high degrees of uniformity as determined by XRD rocking curve. For $NbO_2$ no attempts have yet been made to optimize the growth. Still the uniformity is high enough to consider integrating $NbO_2$ in to neuromorphic devices such as relaxation oscillators or neuristor circuits [15,16].

Nb, $NbO_2$, and $Li_3NbO_4$ all show acceptably low surface roughness values for integration into thin-film heterostructures. BCC niobium films grown at 1000°C show small dense grains and a surface roughness of 1.55 nm RMS. The grain boundary dominated structure is likely due to insufficient substrate temperatures for long range adatom movement because of the refractory nature of niobium. NbO films show the same grainy structure, while $NbO_2$ exhibits small hillocks and a surface roughness of 1.24 nm RMS.

**Table 2.** Thicknesses and symmetric rocking curves determined by XRD of selected single phase films along with RMS roughness determined by AFM. Lower rocking curve values indicate higher uniformity with respect to tilt in the c-axis.

| Material | Growth Rate (nm/h) | Symmetric Rocking Curve (arcseconds) | RMS Roughness (nm) 1 μm² area |
|---|---|---|---|
| Nb | 33 | 180 | 1.55 |
| NbO | 100 | 1192 | 3.23 |
| $NbO_2$ | 430 | 220 | 1.24 |
| $LiNbO_2$ | 85 | 206 | 2.37 |
| $LiNbO_3$ | 57 | 8.6 | * |
| $Li_3NbO_4$ | 55 | 170 | 1.03 |
| $LiNb_3O_8$ | 45 | 810 | 19 |

\* $LiNbO_3$ films are highly columnar with misleading RMS roughness values. Detailed structural and morphological characterization is given in [32].

Tri-lithium niobite, $Li_3NbO_4$ shows a very smooth surface with a roughness of 0.984 nm RMS. This is likely due to the increased metal to oxygen ratio on the surface during growth which acts to increase surface diffusion, smoothing the overall film.

The surface morphology of lithium niobite shows a smooth field with tall, narrow extrusions extending 5-40 nm above the surface of the film. The roughness of the overall surface is 2.37 nm RMS, while the surface of the field as determined by a 1 μm square scan was 0.43 nm RMS. It was found that dipping the $LiNbO_2$ thin film in water and drying under $N_2$ significantly improved the observed surface quality but no surface film attributed spectra were observed in XRD.

The surface roughness of $LiNbO_3$ is dependent on film thickness when grown on sapphire. Films initially show highly columnar non-coalesced grains with atomically flat plateaus, and these grains slowly coalesce as film thickness increases. The surface morphology and crystalline quality of lithium niobate is investigated in greater detail in a previous publication [32].

4. CONCLUSION

A generalized schema is presented for the control of single phase epitaxy in the Li-Nb-O material family using Molecular Beam Epitaxy. It is shown that substrate temperature variations can cause different phases of material to nucleate preferentially over a small temperature range. Combining the empirically determined appropriate growth temperature for the desired phase with the lithium controlled niobium incorporation rate method allows single phase growth of various thin films within the Li-Nb-O material family. Single phase growth of NbO, $NbO_2$, $Li_3NbO_4$, $LiNb_3O_8$, $LiNbO_2$, and $LiNbO_3$ is possible by using this method. Additionally this relationship between flux and temperature has resulted in construction of a phenomenological phase diagram for the $Li-NbCl_5-O_2$ system using MBE at high temperature on c-plane sapphire.

This fundamental understanding will allow new devices and heterostructures to be explored including but not limited to neuromorphic computing elements, strain enhanced sensors and MEMS, high K dielectrics, tunable dielectrics, ferroelectric superlattices, and ferroelectric switching transistors with switchable enhanced channel conductance. With this motivation in mind the materials in the Li-Nb-O family are analyzed by crystal quality and surface morphology for use in future heterostructure devices. In this architecture it is shown that $LiNbO_2$, $LiNbO_3$, and $NbO_2$ are the multi-functional active materials while niobium can be used as a contact layer, subcollector, or waveguide cladding layer, and $LiNbO_3$ or $Li_3NbO_4$ may be chosen as a dielectric.

ACKNOWLEDGMENT

REFERENCES

[1]  A.B. Posadas, A. O'Hara, S. Rangan, R.A. Bartynski, A.A. Demkov, Appl. Phys. Lett. 104 (2014) 9–14. 10.1063/1.4867085.

[2]  A. O'Hara, T.N. Nunley, A.B. Posadas, S. Zollner, A.A. Demkov, J. Appl. Phys. 116 (2014). 10.1063/1.4903067.

[3]  D. Redfield, W.J. Burke, J. Appl. Phys. 45 (1974) 4566–4571. 10.1063/1.1663089.

[4]  A. Dhar, A. Mansingh, J. Appl. Phys. 68 (1990) 5804–5809. 10.1063/1.346951.

[5]  P.P. Sahoo, P. a. Maggard, Inorg. Chem. 52 (2013) 4443–4450. 10.1021/ic302649s.

[6]  Z. Weibin, W. Weidong, W. Xueming, C. Xinlu, Y. Dawei, S. Changle, P. Liping, W. Yuying, B. Li, Surf. Interface Anal. 45 (2013) 1206–1210. 10.1002/sia.5253.

[7]  C. Nico, T. Monteiro, M.P.F. Graça, Prog. Mater. Sci. 80 (2016) 1–37. 10.1016/j.pmatsci.2016.02.001.

[8]  E. Wimmer, R. Podloucky, P. Herzig, a. Neckel, K. Schwarz, Solid State Commun. 42 (1982) iv. 10.1016/0038-1098(82)90178-8.

[9]  R. Weis, T. Gaylord, Appl. Phys. A Mater. Sci. Process. 37 (1985) 191–203. 10.1007/BF00614817.

[10] L.W. Martin, Y.-H. Chu, R. Ramesh, Mater. Sci. Eng. R 68 (2010) 89–133. 10.1016/j.mser.2010.03.001.

[11] J.D. Greenlee, J.C. Shank, M.B. Tellekamp, B.P. Gunning, C.A.M. Fabien, W.A. Doolittle, Cryst. Growth Des. 14 (2014) 2218–2222. 10.1021/cg401775p.

[12] W. Li, C. Hu, M. Zhou, K. Wang, H. Li, S. Cheng, K. Jiang, Electrochim. Acta 189 (2016) 231–236. 10.1016/j.electacta.2015.12.085.

[13] J.C. Shank, M.B. Tellekamp, W.A. Doolittle, J. Appl. Phys. 117 (2015) 35704. 10.1063/1.4906125.

[14] S. Kim, J. Park, J. Woo, C. Cho, W. Lee, J. Shin, G. Choi, S. Park, D. Lee, B.H. Lee, H. Hwang, Microelectron. Eng. 107 (2013) 33–36. 10.1016/j.mee.2013.02.084.

[15] M.D. Pickett, G. Medeiros-Ribeiro, R.S. Williams, Nat. Mater. 12 (2013) 114–7. 10.1038/nmat3510.

[16] S. Li, X. Liu, S.K. Nandi, D.K. Venkatachalam, R.G. Elliman, Appl. Phys. Lett. 106 (2015) 212902. 10.1063/1.4921745.

[17] S.O. Yoon, J.H. Yoon, K.S. Kim, S.H. Shim, Y.K. Pyeon, J. Eur. Ceram. Soc. 26 (2006) 2031–2034. 10.1016/j.jeurceramsoc.2005.09.028.

[18] D. Zhou, H. Wang, L.-X. Pang, X. Yao, X.-G. Wu, J. Am. Ceram. Soc. 91 (2008) 4115–4117. 10.1111/j.1551-2916.2008.02764.x.

[19] M.A. Aegerter, Sol. Energy Mater. Sol. Cells 68 (2001) 401–422. 10.1016/S0927-0248(00)00372-X.

[20] E. a. Mikajlo, K.L. Nixon, V. a. Coleman, M.J. Ford, J. Phys. Condens. Matter 14 (2002) 3587.

[21] N. Yabuuchi, M. Takeuchi, M. Nakayama, H. Shiiba, M. Ogawa, K. Nakayama, T. Ohta, D. Endo, T. Ozaki, T. Inamasu, K. Sato, S. Komaba, Proc. Natl. Acad. Sci. 112 (2015) 7650–7655. 10.1073/pnas.1504901112.

[22] J.C. Shank, M.B. Tellekamp, W.A. Doolittle, in:, 74th Annu. Device Res. Conf. (DRC), Newark, 2016.

[23] Y.J. Hsiao, T.H. Fang, S.J. Lin, J.M. Shieh, L.W. Ji, J. Lumin. 130 (2010) 1863–1865. 10.1016/j.jlumin.2010.04.023.

[24] J. Shank, Memristive Devices for Neuromorphic Computing Applications, Georgia Institute of Technology, 2016.

[25] W.A. Doolittle, A.G. Carver, W. Henderson, J. Vac. Sci. Technol. B Microelectron. Nanom. Struct. 23 (2005) 1272. 10.1116/1.1926294.

[26] W. Doolittle, G. Namkoong, A. Carver, A. Brown, Solid. State. Electron. 47 (2003) 2143–2147. 10.1016/S0038-1101(03)00187-4.

[27] A. Reisman, F. Holtzberg, J. Am. Chem. Soc. 80 (1958) 6503–6507. 10.1021/ja01557a010.

[28] W.E. Henderson, W. Laws Calley, A.G. Carver, H. Chen, W. Alan Doolittle, J. Cryst. Growth 324 (2011) 134–141. 10.1016/j.jcrysgro.2011.03.049.

[29] M.B. Tellekamp, J.D. Greenlee, J.C. Shank, W.A. Doolittle, J. Cryst. Growth 425 (2015) 225–229. 10.1016/j.jcrysgro.2015.03.042.

[30] J.C. Shank, M. Brooks Tellekamp, W. Alan Doolittle, Thin Solid Films 609 (2016) 6–11. 10.1016/j.tsf.2016.01.030.

[31] K. Kitamura, Y. Furukawa, S. Takekawa, S. Kimura, SINGLE CRYSTAL OF LITHIUM NIOBATE OR TANTALATE AND ITS OPTICAL ELEMENT, AND PROCESS AND APPARATUS FOR PRODUCING AN OXIDE SINGLE CRYSTAL, US 6,464,777 B2, 2002. 10.1016/j.(73).

[32] M.B. Tellekamp, J.C. Shank, M.S. Goorsky, W.A. Doolittle, J. Electron. Mater. (2016). 10.1007/s11664-016-4986-3.